\documentclass[prl,twocolumn,showpacs,amsmath,amssymb,aps,superscriptaddress]{revtex4-1}
\usepackage[dvips]{graphicx}
\usepackage[hypertex]{hyperref}
\newcommand{\tEWS}{t_{\scriptstyle\textrm{EWS}}}
\newcommand{\tCLC}{t_{\scriptstyle\textrm{CLC}}}
\newcommand{\tdLC}{t_{\scriptstyle\textrm{dLC}}}
\newcommand{\tS}{t_{\scriptstyle\textrm{S}}}

\begin{document}

\title{Attosecond streaking of Cohen-Fano interferences in the photoionization of H$_2^+$}

\author{Qi-Cheng Ning}
\affiliation{State Key Laboratory for Mesoscopic Physics and Department of Physics, Peking
University, Beijing 100871, China}

\author {Liang-You Peng}
\email{liangyou.peng@pku.edu.cn}
\affiliation{State Key Laboratory for Mesoscopic Physics and Department of Physics, Peking
University, Beijing 100871, China}
\affiliation{Collaborative Innovation Center of Quantum Matter, Beijing, China}

\author{Shu-Na Song}
\affiliation{State Key Laboratory for Mesoscopic Physics and Department of Physics, Peking
University, Beijing 100871, China}

\author{Wei-Chao Jiang}
\affiliation{State Key Laboratory for Mesoscopic Physics and Department of Physics, Peking
University, Beijing 100871, China}

\author{Stefan Nagele}
\email{stefan.nagele@tuwien.ac.at}
\affiliation{Institute for Theoretical Physics, Vienna University of Technology, 1040 Vienna, Austria, EU}

\author{Renate Pazourek}
\affiliation{Institute for Theoretical Physics, Vienna University of Technology, 1040 Vienna, Austria, EU}

\author{Joachim Burgd\"orfer}
\affiliation{Institute for Theoretical Physics, Vienna University of Technology, 1040 Vienna, Austria, EU}

\author {Qihuang Gong}
\email{qhgong@pku.edu.cn}
\affiliation{State Key Laboratory for Mesoscopic Physics and Department of Physics, Peking
University, Beijing 100871, China}
\affiliation{Collaborative Innovation Center of Quantum Matter, Beijing, China}

\date{\today}

\begin{abstract}

We present the first numerical simulation of the time delay in the photoionization of the simplest diatomic molecule H$_2^+$ as observed by attosecond streaking.
We show that the strong variation of the Eisenbud-Wigner-Smith time delay $\tEWS$ as a function of energy and emission angle becomes observable in the streaking time shift $\tS$ provided laser field-induced components are accounted for.
The strongly enhanced photoemission time shifts are traced to destructive Cohen-Fano (or two-center) interferences.
Signatures of these interferences in the streaking trace are shown to be enhanced when the ionic fragments are detected in coincidence.

\end{abstract}

\maketitle

Since the very first experimental demonstration of attosecond pulses and pulse trains~\cite{Paul,Hentschel}, these new light sources, synchronized with an infrared~(IR) laser pulse, have enabled the probe and control of the electronic dynamics inside atoms and molecules on its natural time scale (see e.g.\ \cite{Kling, Krausz} and references therein).
In particular, the attosecond streak camera \cite{Kienberger, Cavalieri, Schultze, Itatani, Kitzler} and interferometric approaches including the RABBITT~(Reconstruction of Attosecond Beating By Interference of Two-photon Transitions) technique \cite{Schafer,Remetter,Swoboda,Johnsson,Johnsson2007,Klunder,Guenot} have been developed into powerful tools to access the quantum phase information and phase variation in atomic and molecular ionization. With these advances, attosecond physics now allows one to address fundamental quantum mechanical issues in the time domain, in particular, how long it takes for an electron to escape from the atomic core  or a solid surface following  absorption of a photon~\cite{Hart}.

First proof-of-principle experimental demonstrations include measurements of time-delayed photoemission from the core levels relative to conduction band states of a tungsten surface~\cite{Cavalieri} and more recently, a time delay of 21$\pm 5\,$as between photoemission from 2$p$ and 2$s$ subshells of atomic neon~\cite{Schultze}.
These observations have triggered a large number of theoretical investigations of the time delay in atomic photoionization \cite{Zhang,Kheifets10,Nagele,Nagele_PRA,Pazourek,Pazourek_FD,Madsen,Misha,Kheifets13,Dahlstrom,Dahlstrom12b,Dahlstrom2013,Carette,Spiewanowski,Su,Su_2,Su_3,Ivanov13,Moore,Carette,Dixit}, molecular photoionization \cite{Serov, Ivanov12, Dixit}, and photoemission from surfaces \cite{Kazansky, Thumm, Lemell}.
For example, the methods employed to describe the photoionization process in neon include
the state-specific expansion approach \cite{Schultze},
the random phase approximation with exchange~(RPAE)~\cite{Kheifets10,Kheifets13}, diagrammatic many-body perturbation theory~\cite{Dahlstrom12b}, the time-dependent {\it R}-matrix method~\cite{Moore}, and the {\it B}-spline {\it R}-matrix method \cite{Feist}, none of which are, however, up to now able to account for the experimentally observed delay.
Similarly, the difference in  photoionization time delays between electrons emitted from 3$s^2$ and 3$p^6$ shell of Ar were measured with the RABBITT technique \cite{Klunder,Guenot} and yield for photon energies close to the Cooper minimum~\cite{Cooper,Fano} sizable discrepancies to calculations using the RPAE method \cite{Guenot, Kheifets13}, the multiconfigurational Hartree-Fock close-coupling ansatz \cite{Carette}, or the time-dependent local density approximation \cite{Dixit}.
The origin of these discrepancies for such a fundamental property of atomic photoemission remains a widely open question with so far unaccounted multi-electron correlations being one of the prime suspects.
Indeed, recent {\it ab initio} simulations on one-electron attosecond streaking of He  with the shakeup of the second electron clearly demonstrate the significant contribution of electron-electron correlation to the photoionization time delays~\cite{Pazourek}.

Time-resolved photoionization of more complex systems with internal geometric structure promises novel insights into the formation of an outgoing wave packet emerging from the complex. The simplest prototypical case is the photoionization of a diatomic molecule \cite{Serov,Martin}.
The fundamental questions to be addressed are: does it take a longer time for the electron to escape from the multi-center molecular core than from the one-center atomic core, does the emission time delay dependence on the energy and on the relative orientation of the emission direction and molecular axis carry information on the geometric arrangement of the atomic constituents, and, most importantly, are those effects observable in an attosecond streaking setting.
We present in the following an {\it ab initio} simulation of the attosecond streaking time delay $\tS$ and the extraction of the intrinsic Eisenbud-Wigner-Smith time delay $\tEWS$ \cite{Eisenbud,Wigner,Smith} for H$_2^+$ in a few-cycle infrared laser field.
We show that the Cohen-Fano (or two-center) interferences \cite{Cohen,Zuo} leave a clear and observable mark on the streaking time delay.
Destructive path interferences suppress the formation of the electronic wave packet and greatly increase the magnitude of the EWS time delay.

In our numerical attosecond streaking simulation,
the time-dependent electronic Schr\"odinger equation~(TDSE) of H$_2^+$ in the presence of an ionizing extreme ultraviolet (XUV) attosecond pulse and an infrared (IR) streaking field is accurately solved in prolate spherical coordinates.
The momentum spectrum of the ionized electron is extracted by projecting the wave packet after the end of the laser pulses onto the molecular scattering states $\psi _f^ -$ with incoming wave boundary conditions (see \cite{Hou} and references therein for details).
The photoionization is calculated at fixed internuclear distance $R$ and fixed orientation of the internuclear axis as a vertical Franck-Condon-like electronic transition between Born-Oppenheimer (BO) potential surfaces.
This approximation is well justified for ultrashort attosecond XUV pulses.
The distribution of $R$ in the vibronic ground state as well as the distribution of the alignment angle of laser-aligned molecules is taken into account through an ensemble average over Franck-Condon transitions.
For comparison, also simulations for atomic targets are performed within the single-active-electron (SAE) approximation \cite{Xu,Frolov}.
All results were checked for convergence by variation of the temporal and spatial parameters in the TDSE simulations.

\begin{figure}
\includegraphics[width=0.95\linewidth, angle=270]{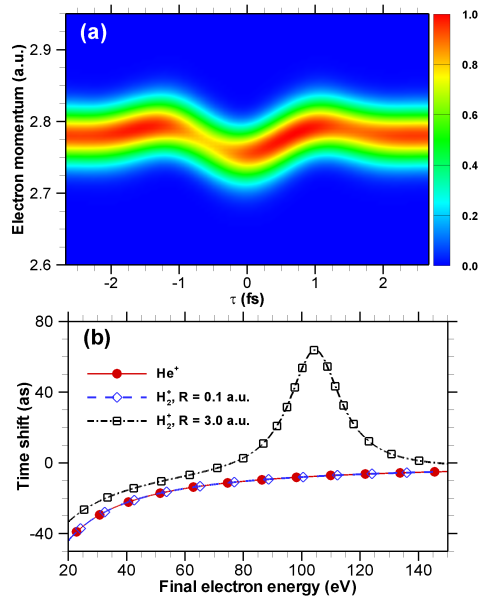}
\caption{(color online). (a) Typical streaking spectrogram for ionization of H$_2^+$ by a 130~eV XUV pulse [$\theta_x=0^\circ$, $\tau_\mathrm{XUV}=400$~as (FWHM)] probed by a 800nm two-cycle streaking field with intensity $I_\text{IR} = 10^{11}$ W/cm$^2$ for electron emission along the polarization axis of the streaking field ($\theta_e=0^\circ$) and $R=3\,$a.u.
(b) Extracted streaking delays $\tS$ for He$^+$ (red circles), H$_2^+$ at $R=0.1\,$a.u. (blue diamonds), and H$_2^+$ at $R=3.0\,$a.u. (black squares) for pulse parameters $\tau_\mathrm{XUV}=600$~as (FWHM) and $I_\text{IR} = 10^{8}$~W/cm$^2$.}
\label{Fig1}
\end{figure}

\begin{figure}
\includegraphics[width=0.5\linewidth, angle=270]{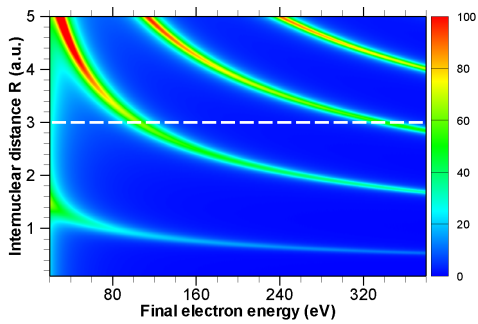}
\caption{(color online). EWS delay $\tEWS$ in the $(E,R)$ plane for emission along the internuclear axis $\theta_e=0^\circ$ and the XUV pulse polarized parallel  to the internuclear axis, $\theta_x=0^\circ$.
The EWS delays that are contained in the streaking results of Fig.~\ref{Fig1}(b) [Eq.\ (\ref{eq:delays})] correspond to cuts for constant $R$ (dashed line for $R=3$).}
\label{Fig2}
\end{figure}

The XUV pulses of mean frequency $\omega_\mathrm{XUV}$ employed in the streaking simulations have a Gaussian envelope with a full width at half maximum~(FWHM) duration ranging from $\tau_\mathrm{XUV} = 400\,$as to $\tau_\mathrm{XUV} = 600\,$as, a peak intensity of $I_\text{XUV} = 10^{12}$ W/cm$^2$, and are polarized parallel to the internuclear axis, $\theta_x=0^\circ$.
The comparatively long $\tau_\mathrm{XUV}$ is chosen to reduce bandwidth effects which would obscure sharp spectral features.
The streaking IR field has a duration of two optical cycles at a wavelength of 800~nm with a cosine-squared envelope.
In order to minimize distortion effects of the initial states as well as rescattering effects \cite{Hou,Xu} we consider weak IR streaking fields ranging in intensity $I_\text{IR}$ from $10^{8}$ to $10^{11}$ W/cm$^2$ and check for the independence of the extracted time information of $I_\text{IR}$.
The total time delays $t_\text{S}$  measured relative to the peak of the XUV pulse are extracted by fitting the first moment of the electron momentum along the laser polarization~($\theta_e = 0^\circ$) to the IR vector potential~\cite{Nagele} [Fig.~\ref{Fig1}(a)].
When the internuclear distance $R$ is very small~[$R$ = 0.1 a.u., Fig.~\ref{Fig1}(b)],  the streaking delays of H$_2^+$ and the isoelectric atomic partner He$^+$ nearly coincide rendering molecular effects on the time delay entirely negligible.
At larger distances, e.g.~$R=3$ [Fig.~\ref{Fig1}(b)], a remarkably different picture emerges: the streaking delay is strongly enhanced with a broad maximum reaching values $\tS \approx 64\,$as.
This signature of pronounced molecular structures in the streaking time delay can be traced to the behavior of the intrinsic Eisenbud-Wigner-Smith (EWS) time delay, $\tEWS$ (see Fig.~\ref{Fig2}).
For atomic photoionization the additivity of the EWS delay and the IR-field induced Coulomb-laser coupling (CLC) and dipole-laser coupling (dLC) delays \cite{Nagele, Pazourek, Pazourek_FD}
\begin{equation}\label{eq:delays}
\tS =\tEWS + \tCLC + \tdLC
\end{equation}
has been established.
Since the long-range Coulomb interaction ($\tCLC$) and dipole potentials are independent of short-ranged potentials, Eq.\ (\ref{eq:delays}) is expected to hold for molecules as well (for non-polar molecules $\tdLC = 0$).
The EWS time delay $t_\text{EWS}$ for the H$_2^+$ molecule can be calculated (separately from the TDSE solution) from the energy derivative of the phase of the exact dipole transition element in the absence of a streaking field
\begin{equation}
t_\text{EWS}(E, R, \theta_e, \theta_x) = \frac{\partial}{\partial E}\arg \left(\langle \psi _f^- (E, R, \theta_e)  |\vec d \cdot \hat e| \phi _0 \rangle \right),
\label{EWS}
\end{equation}
where $E$ is the final continuum energy of the electron and $\theta _e$ is the electron ejection angle relative to the internuclear axis.
The angle of the polarization axis of the XUV field, $\hat e$, relative to the internuclear axis is denoted by $\theta_x$.

The landscape of the EWS delay as a function of the fixed internuclear distance $R$ and the emission energy $E=\hbar \omega_\mathrm{XUV} - I_p$ (Fig.~\ref{Fig2}) reveals that the peak in $\tS$  (Fig.~\ref{Fig1}) directly maps out the peak in the EWS time distribution.
The residual difference between $\tS$ and $\tEWS$ (not shown) is given by the CLC delay $\tCLC$ as predicted by Eq. (\ref{eq:delays}),
validating Eq. (\ref{eq:delays}) also for molecules. In particular, the CLC contributions $\tCLC$ for the ionization of the H$_2^+$ molecule and its isoelectric atomic partner He$^+$ \cite{Nagele} are equivalent.

\begin{figure}
\includegraphics[width=1.1\linewidth, angle=270]{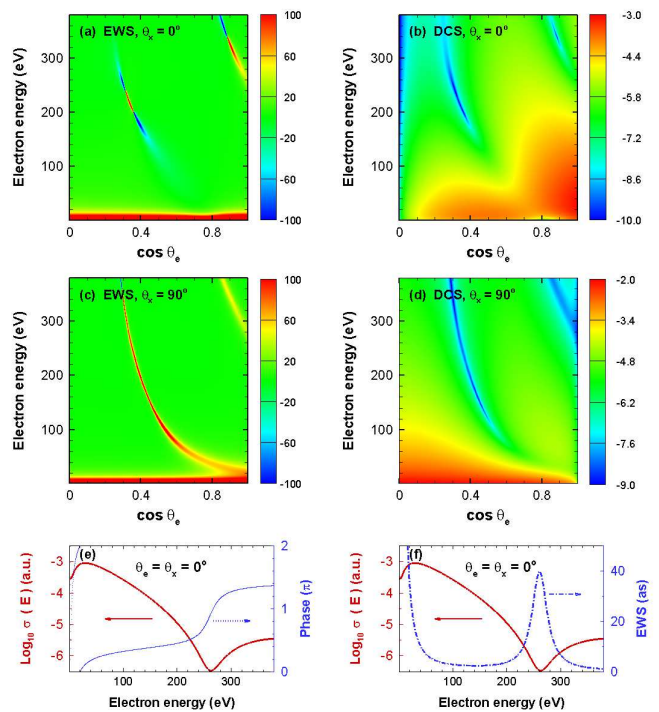}
\caption{(color online). EWS time delays of H$_2^+$ at the equilibrium distance $R$ = 2 a.u., plotted as a function of electron energy $E$ and ejection angle $\theta _e$ [$t_\text{EWS}\left(E, \theta _e\right)$] for the XUV pulse polarized (a) parallel ($\theta_x$ = 0$^\circ$) or (c) perpendicular ($\theta_x$ = 90$^\circ$) with respect to the molecular axis. The corresponding single-photon differential ionization cross sections (DCS) in (b) and (d) are plotted on a logarithmic scale. A cut of the DCS at $\theta_e=0$  for the case $\theta_x=0$ is plotted along with the phase of the transition matrix element in (e) and along with the EWS delay $t_\text{EWS}$ in (f).}
\label{Fig3}
\end{figure}

\begin{figure}
\includegraphics[width=0.7\linewidth, angle=0]{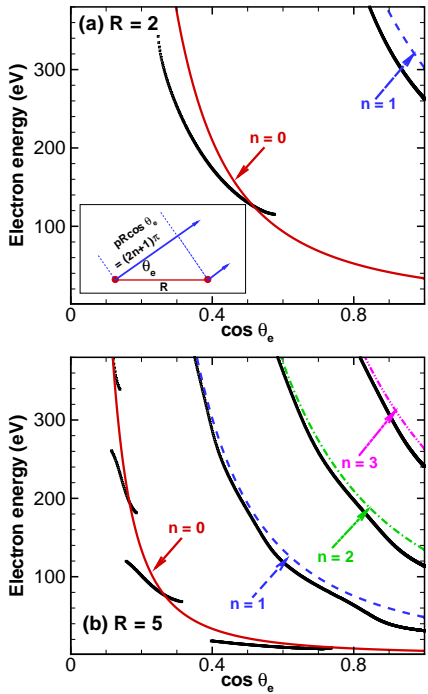}
\caption{(color online). Comparison between the location of the minima in the differential photoionization cross section (DCS) in the $(E,\cos\theta_e)$ plane with the prediction of the two-center interference model according to $\sqrt{2E}R\text{cos}\theta_e = (2n+1)\pi$ [see inset in (a)] for different $n$. DCS is  for a Franck-Condon transition at $R=2$ (a) and $R=5$ (b).}
\label{Fig4}
\end{figure}

The distribution of $\tEWS$ in the $\theta_e$-$E$ plane (Fig.~\ref{Fig3}) shows that the ridges of enhanced delays are also directly linked to minima in the energy- and angle differential cross section (DCS).
Their contour lines $E(\cos\theta_e)$ are largely independent of the polarization of the ionizing pulse pointing to a continuum final-state effect.
The minima in the DCS with the accompanying maxima in the EWS delays are reminiscent of the behavior of Cooper minima \cite{Cooper, Fano} characterized by a rapid phase jump by $\pm \pi$ of the transition matrix element [Fig.~\ref{Fig3}(e)].
In the present case, however, they allow for a simple intuitive description in terms of a semiclassical path interference.
Cohen and Fano \cite{Cohen} pointed out that the energy and angular distribution of the photoelectrons from diatomic molecules feature two-center path interferences.
Destructive interference between emission from the two centers occurs when the electron momentum $\bf p$ and the internuclear distance vector $\bf R$ satisfy $\bf {p} \cdot \bf {R}$ = $pR\text{cos}\theta_e = (2n+1)\pi$, valid in the limit of high electron energies when the outgoing waves can be approximated by plane waves (Fig.~\ref{Fig2}).
The lines in the $\cos\theta_e$-$E$ plane for which destructive interference $\sqrt{2E}R\text{cos}\theta_e = (2n+1)\pi$ holds, approximates the minima in the DCS and, in turn, the extrema in $\tEWS$ quite well (Fig.~\ref{Fig4}) confirming two-center interferences as the origin of the enhanced EWS time shifts.
Taking into account the phase distortion due to two-center Coulomb effects within an eikonal approximation is expected to further improve the agreement.
The destructive interference suppresses the emission of the outgoing wave packet and the magnitude of the corresponding $\tEWS$ delay is significantly increased.

It is now important to inquire into the observability of this two-center interference enhanced time delay in a realistic streaking setting.
As the destructive interference is associated with a minimum in the cross section, visibility in the streaking signal is a priori not obvious.
For the same reason, the enhanced time delay near Cooper minima has, up to now, escaped detection.
We therefore include in our calculations corrections beyond the Franck-Condon transitions for photoemission from the molecule at the (fixed) equilibrium internuclear distance and beyond space-fixed orientation of the internuclear axis.
The vibrational ground state of the H$_2^+$ molecule is described by a probability density $W_0(R)$ centered at the equilibrium distance $R_0 = 2$ a.u. More generally, we average the observable $O(R)$ calculated in the BO limit over an $R$-distribution $W(R)$ weighted by the corresponding ionization cross-section $\sigma(R)$,
\begin{equation}
\left\langle O \right\rangle _R = \frac{{\int {O\!\left(R\right)W\!\left(R\right)\sigma(R)dR} }}{\int {W\!\left(R\right)\sigma(R)dR} }.
\label{vibration}
\end{equation}
Similarly, we take into account that for a laser-aligned molecule the molecular axis will have an angular distribution around the alignment axis for which we assume a Gaussian distribution with full width at half maximum of up to $15^\circ$.
\begin{figure}
\includegraphics[width=0.7\linewidth, angle=0]{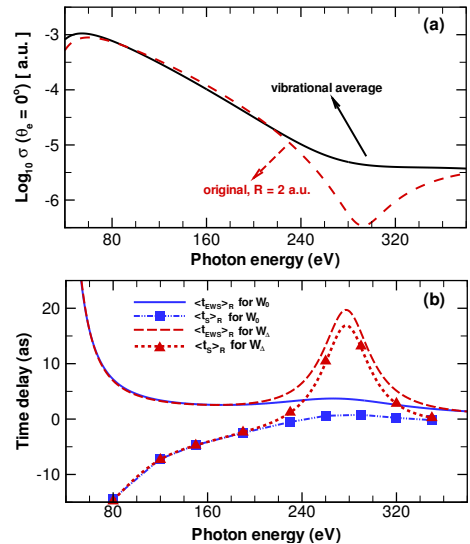}
\caption{(color online). (a) Comparison between the differential cross section (DCS) at fixed internuclear equilibrium distance $R=2$ and averaged over the vibrational ground state distribution $W_0$. (b) Expectation values $\langle\tEWS\rangle_R$ and $\langle\tS\rangle_R$ after averaging over the vibrational ground state distribution $W_0$ (squares) and the distribution $W_\Delta$ (triangles) postselected from the Coulomb explosion of the molecular fragments. The XUV pulse duration in the streaking simulations is $\tau_\mathrm{XUV}=600$~as (FWHM) and the intensity of the probing 800~nm field is $I_\text{IR} = 10^{8}$~W/cm$^2$.}
\label{Fig5}
\end{figure}
Averaging over an alignment angle of $15^\circ$ hardly influences the resulting EWS delays (not shown) and, thus, has negligible effects on the streaking delays.
On the other hand, the vibrational average over $W_0(R)$ fills in the deep minimum in the photoionization cross section caused by the two-center interference [Fig.~\ref{Fig5}(a)].
As a result, the prominent peak in the EWS delay disappears and is reduced to a very broad ridge on the 1 to 2 attosecond level in $\left\langle \tEWS \right\rangle_R$, and, in turn, in the averaged streaking delay $\left\langle \tS \right\rangle_R$ [Fig.~\ref{Fig5}(b)].
However, since photoionization of H$_2^+$ initiates the Coulomb explosion of the ionic fragments  \cite{Frasinski,Chelkowski}, energy resolved coincident detection of one outgoing proton allows to experimentally postselect a narrow $R$-distribution $W_\Delta(R)$ within the ground state vibrational distribution.
This additional ``knob'' allows to enhance the interference contrast in the time shifts by reducing the vibrational averaging.
Coincident detection of a proton near the Coulomb explosion energy corresponding to the equilibrium distance, $E_\mathrm{kin} = 1/2R_0$,
with an energy resolution (FWHM) of $0.5$~eV selects a narrow radial distribution $W_\Delta(R)$ centered at $R_0$ with a width of $\Delta R = 0.15$~a.u.
The reduced vibrational average, Eq.\ (\ref{vibration}), now yields clearly visible peaks in the EWS and streaking time delays of the order of 10~as [Fig.~\ref{Fig5}(b)]
as signatures of the destructive interference.

For the present streaking simulation we have used a very low IR intensity of $10^{8}$ W/cm$^2$ in order to preclude any polarization effects due to a strong IR field which might play, unlike for streaking of a structureless electron spectrum \cite{Nagele}, a more important role near zeros of the photoionization cross section.
Indeed, increasing the streaking intensity to a typical experimental value of $10^{11}$ W/cm$^2$, the streaking trace becomes forward-backward ($\theta_e=0^\circ$ relative to $\theta_e=180^\circ$) asymmetric and the extracted $\tS$ differ from each other, indicating the influence of IR multi-photon interferences.
Remarkably, the average over the forward-backward asymmetry, $[\tS(\theta_e=0^\circ)+\tS(\theta_e=180^\circ)]/2$, agrees very well with $\tS$ extracted at low intensity thereby cancelling out IR polarization effects (or multi-photon interferences) and allows for an intensity independent measurement of $\tS$.
We are therefore confident that the predicted enhanced streaking delays due to two-center interferences are experimentally accessible.

Summarizing, we have simulated the attosecond streaking of photoionization of the simplest molecule H$_2^+$ by solving the time-dependent Schr\"odinger equation in the presence of the ionizing XUV and streaking IR fields.
Our calculations include the effects of the vibrational distribution and the distribution of the molecular alignment angle.
We have shown that from the observable streaking delay the intrinsic Eisenbud-Wigner-Smith time delay can be unambiguously extracted.
Formation of the wave packet of the ionized electron is significantly modified when the Cohen-Fano condition for destructive two-center interference is met.
We hope the current study of H$_2^+$ will stimulate experiments on photoionization time delays of molecules.

{\it Acknowledgment} LYP appreciates useful discussions and suggestions of Lars Bojer Madsen at the initial stage of this work.
This work is supported  in part by the National Natural Science
Foundation of China under Grant No.~11322437, by  the 973 Program under Grant No.~2013CB922402, by the Program for New Century Excellent Talents in University~(NCET), and by the FWF (Austria) SFB041 (VICOM) and SFB049 (NEXTLITE).
The computational results presented have been achieved in part by using the computer cluster ``MESO" in the State Key Laboratory for Mesoscopic Physics at Peking University, by using the Vienna Scientific Cluster (VSC), and through XSEDE resources provided under Grant TG-PHY090031.

\end{document}